# GROUND-BASED DIRECT DETECTION OF EXOPLANETS WITH THE GEMINI PLANET IMAGER (GPI)

James R. Graham[1], Bruce Macintosh[2], Rene Doyon[3], Don Gavel[4], James Larkin[5], Marty Levine[6], Ben Oppenheimer[7], David Palmer[2], Les Saddlemyer[8], Anand Sivaramakrishnan[7], Jean-Pierre Veran[8], & Kent Wallace[6]

## *Abstract*

The Gemini Planet (GPI) imager is an "extreme" adaptive optics system being designed and built for the Gemini Observatory. GPI combines precise and accurate wavefront control, diffraction suppression, and a speckle-suppressing science camera with integral field and polarimetry capabilities. GPI's primary science goal is the direct detection and characterization of young, Jovian-mass exoplanets. For systems younger than 2 Gyr exoplanets more massive than 6 $M_J$ and semimajor axes beyond 10 AU are detected with completeness greater than 50%. GPI will also discover faint debris disks, explore icy moons and minor planets in the solar system, reveal high dynamic range main-sequence binaries, and study mass loss from evolved stars. This white paper explains the role of GPI in exoplanet discovery and characterization and summarizes our recommendations to the NSF-NASA-DOE Astronomy and Astrophysics Advisory Committee ExoPlanet Task Force.

[1] UC Berkeley
[2] Lawrence Livermore National Lab
[3] Université de Montréal
[4] UC Santa Cruz
[5] UCLA
[6] JPL
[7] American Museum of Natural History
[8] NRC/Herzberg Institute

## Introduction

Direct detection affords access to exoplanet atmospheres, which yields fundamental information including effective temperature, gravity, and composition. Mid-IR exoplanetary light has been detected in secondary eclipses by *SPITZER* (Grillmair et al. 2007; Richardson et al. 2007). However, such information is necessarily limited by the overwhelming photon shot noise of the primary. Direct detection separates the exoplanet light from that of its primary so that this noise source is suppressed.

When the fundamental atmospheric properties are known for a population of planets spanning a range of mass and age it will be possible to conduct critical tests of formation scenarios and evolutionary models. The goals of direct detection include measurement of abundances and reconstruction of the thermal history. Such information will discriminate between idealized adiabatic contraction (Burrows et al. 2004), more realistic accretion models that occurs in core-accretion (Marley et al. 2007), and formation by gravitational instability (Boss 2006). Ultimately, the impact of composition and equation of state on evolution can also be investigated.

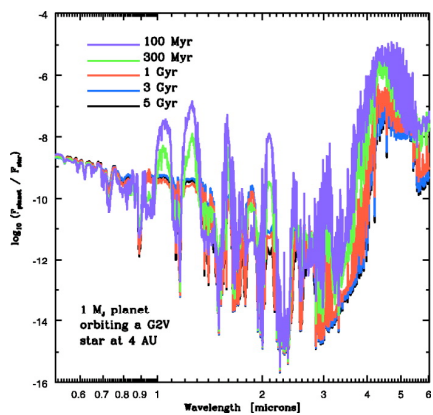

Figure 1: Planet-to-star contrast for a 1 $M_J$ exoplanet orbiting a G2 V star at 4 AU for ages 0.1–5 Gyr (Burrows et al. 2004). Planets shine by reflected light in the visible. Young exoplanets (< 1 Gyr for 1 $M_J$) are detectable by their intrinsic luminosity. Note strong *J* (1.25 $\mu$m) and *H* (1.65 $\mu$m) emission where ground-based facilities have sensitive operation. Old exoplanets (> 1 Gyr) shine only by reflected light at short wavelengths (< 2 $\mu$m).

Direct detection has several practical advantages. Planets at large semimajor axes can be found without waiting for an orbit to complete—a condition that renders indirect detection of Neptune like planets (*a* = 30 AU, *P* = 160 yr) impractical. The Fourier decomposition that underlies Doppler and astrometric detection are also subject to aliasing and beat phenomena, and suffer confusion when multiple planets are present. Direct imaging gives an immediate portrait of a planetary system, including sensitivity to the analogs of Zodiacal and Kuiper dust belts.

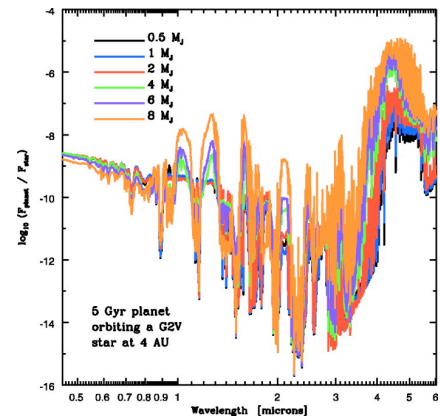

Figure 2: Contrast for a 5 Gyr system orbiting a G2 V star at 4 AU. Masses are 0.5–8 $M_J$. Massive, solar-system age exoplanets shine only by reflected light at visible wavelengths, but remain sufficiently self-luminous to be detectable at contrast ratios > 3 x $10^{-8}$ in the near-IR.

Surveys at large semimajor axes are necessary for several fundamental reasons. The structure of our own solar system implies that a full picture of planet formation cannot be constructed without reaching out to 30 or 40 AU. Moreover, if planets form as a consequence of disk instabilities, they are most likely to occur at large semimajor axis separation (> 20 AU for solar type stars) where the cooling time exceeds the Keplerian shear time scale.

Indirect searches continue to hint that the semimajor axis distribution is at least flat or even rising in d$N$/dlog($a$) beyond 5 AU. With only 200 Doppler planets known there are too few to make a statistically significant study of trends with planetary and stellar mass, semimajor axis, and eccentricity. Because Doppler methods have reached the domain of diminishing returns ($a = P^{2/3}$), improving the statistics will be challenging. A direct search that can efficiently tap into the dominant population of giant exoplanets beyond 5 AU can bring statistical significance to these studies.

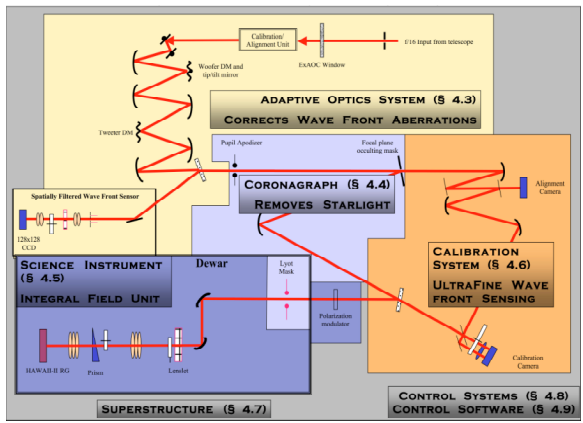

Figure 3: A schematic of the Gemini Planet Imager showing the major subsystems.

## Detecting Self-Luminous Exoplanets

Direct detection of Jovian-mass planets via their reflected sunlight requires a contrast ratio of order $2\times10^{-9}(a/5AU)^{-2}$ relative to their parent star. Because of the inverse square law, reflected light searches are an impractical way to explore the outer regions of solar systems. GPI seeks to detect the energy radiated by the planet itself, which is independent of $a$, except for very small semimajor axis separations. Old planets are cool and dim, but young planets are hot and therefore bright in the infrared relative to their parent star (Figure 1 & Figure 2). For example, at 1.6 $\mu$m it is possible to detect a 10 Myr-old 3 $M_J$ planet, or a 100 Myr-old 7 $M_J$ planet, orbiting a G2V star at a contrast ratio of only $4 \times 10^{-6}$. With improved contrast, an increasingly large phase space of planets becomes accessible. Better contrast is obviously preferable, but it comes at a penalty. For example, Angel (1994) described an AO system designed to achieve very high contrast ratios using bright guide stars. This system would be suitable for exploring the planetary systems of 13 bright stars ($R < 3.8$ mag.) in the solar neighborhood (< 8 pc). However, the detection of a few planets, although dramatic, would be insufficient scientific impetus given the success of Doppler searches. It is necessary to show that any proposed instrument can search a scientifically interesting range of semimajor axes and accumulates a statistically significant sample of exoplanets in reasonable observing time.

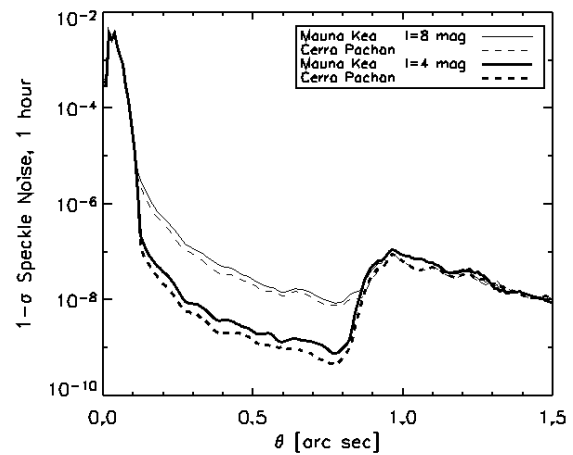

Figure 4: The 1-$\sigma$, 1-hour speckle noise at $H$ for GPI with 18-cm subapertures and a maximum update rate of 2.5 kHz. The speckle noise is measured in units of the guide star brightness. Therefore, these curves represent the achievable contrast when speckle noise is dominant. The system performance is a function of guide star magnitude (thick vs. thin lines). The Fried parameter (at 500 nm) for Mauna Kea is 18.7 cm and 14.5 cm for Cerro Pachon. The coronagraph occulter is opaque within 0.09 arc seconds.

### Overview of GPI

The Gemini Planet Imager (GPI) is a high contrast or "extreme" AO (ExAO) system (Macintosh et al. 2006). Basic R&D enabling the construction of a practical system was supported by the NSF's Center for Adaptive Optics. In 2004, Gemini supported two ExAO conceptual design studies as part of its "Aspen Process" next generation instrumentation program. At the conclusion of those studies, the Gemini Board recently selected GPI to proceed. Work began in June 2006 and delivery of the instrument is expected in late 2010.

Imaging a Jovian exoplanet requires a contrast significantly better than that delivered by existing astronomical adaptive optics (AO) systems. Currently achievable contrast, about $10^{-5}$, is completely limited by quasi-static wavefront errors, so that contrast does not improve with integration times longer than about 1 minute. Moreover, there are enough slow drifts in these errors that PSF subtraction does not increase contrast by more than a factor of a few, except in the most ideal circumstances. GPI will surpass the performance of existing systems by two orders of magnitude.

Table 1: Principal properties of GPI

| Adaptive Optics | | Calibration WFS | |
|---|---|---|---|
| Deformable mirror | 4096-actuator Boston Micro-machines MEMS | Type | 1–2.4 $\mu$m interferometer |
| Subaperture size | 18 cm ($N$ = 44 subaps.) | Static WFE Accuracy | 1 nm RMS |
| Limiting mag. | $I = 8$ mag. (goal: $I = 9$ mag.) | **Science Instrument** | |
| Optics quality | < 5 nm RMS WFE per optic | Type | Lenslet-based integral field unit |
| **Coronagraph** | | Spatial sampling | 0.014 arc seconds per lenslet |
| Type | Apodized-pupil Lyot coronagraph | Field of view | 2.8 arc second square |
| Inner Working Dist. | ~ 3 $\lambda/D$ | Spectral coverage | $Y$, $J$, $H$ or $K$ |
| Throughput | 60% | Spectral resolution | $\lambda/\Delta\lambda \approx 45$ |

To achieve our science goals, GPI integrates four key subsystems (Figure 3): 1) An AO system that makes fast measurement of the instantaneous wavefront, and provides wavefront control via deformable mirrors; 2) A calibration unit that provides precise and accurate measurements of the time-averaged wavefront at the science wavelength, so that the final image is not dominated by persistent speckles caused by quasi-static wavefront errors; 3) A coronagraph that controls diffraction and pinned speckles; and 4) An integral field spectrograph unit (IFU) that records the scientific data, providing low-resolution spectroscopy and suppression of residual speckle noise. The IFU incorporates a dual-channel polarimeter for studies of circumstellar dust. An example of GPI contrast performance based on detailed simulation of this system is shown in Figure 4.

Technical aspects of GPI have been presented in the recent literature, including: apodized pupil Lyot coronagraphs (Soummer 2005, Sivaramakrishnan & Lloyd 2005); optimal wavefront control for ExAO (Poyneer et al. 2006); MEMS deformable mirrors (Morzinski et al. 2006); simultaneous spectral differential imaging and the effects of out-of-pupil-plane optical aberrations (Marois et al. 2006); multiwavelength speckle noise suppression (Marois et al. 2004); and the wavefront calibration system (Wallace et al. 2006).

## GPI's ExoPlanet Parameter Space

The performance of GPI is characterized by the detectable brightness ratio. The achievable contrast will be a function of the brightness of the wavefront reference, the angular separation, and observing and wavefront-sensing wavelengths. The detectability of planets in a given sample of target stars can then be estimated by comparing the distribution of relative exoplanet brightness versus angular separation with the expected performance. This comparison also quantifies selection effects that vary with properties of the planet (age, mass, and orbital elements) and of the primary star (spectral type and distance).

Our knowledge of the distribution of planetary properties is incomplete, but a basic premise is that sufficient information exists to make a preliminary estimate of this distribution. Given these predictions it is possible to estimate the scientific impact of different design choices, e.g., precision and accuracy of adaptive optics correction and observing wavelength, and therefore make an informed trade-off between cost and performance.

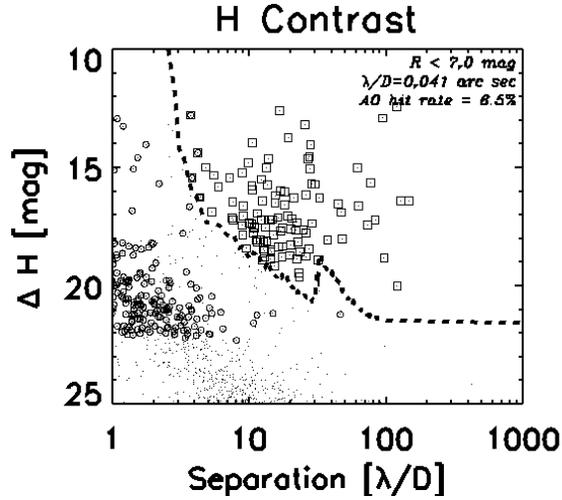

Figure 5: Detectable companion contrast versus angular separation for GPI, showing the direct detection of young luminous planets in a hypothetical survey of field (< 50 pc) stars. The small dots represent the planet population: those detected by GPI are drawn with a box, those detectable in current Doppler surveys are shown with a circle. The dashed line shows the GPI contrast threshold (5 σ) for a 1-hour exposure at 1.65 $\mu$m. Within 100 $\lambda$/D speckle noise dominates. In this example, which has no speckle noise suppression and a fixed frame rate and simple AO model, a universal contrast curve applies to all targets.

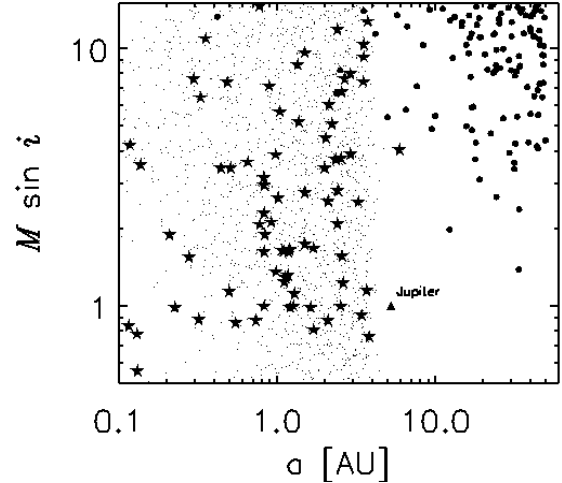

Figure 6: Exoplanets from the simulated GPI field star survey. Heavy filled circles are GPI detected planets from. Light dots are planets detected by a hypothetical 8-year astrometric interferometer survey, with a limit of $R$ < 10 mag., and a precision of 30 $\mu$as. Exoplanets detected in the Keck/Lick Doppler survey are shown as stars. This illustrates how GPI explores a complementary phase space to indirect searches.

The approach we adopt involves making a Monte Carlo model for the population of planets in the solar neighborhood (Graham et al. 2002). This description includes the mass, age and orbital elements of each accompanying planet. When combined with cooling curves, model atmospheres and the distance to the primary star we can compute the brightness ratio and angular separation at any epoch. Our adopted planet properties are based on the precision Doppler monitoring of about 2000 F, G and K stars by eight exoplanets searches at the AAT, Lick, Keck, ESO/Coralie, Provence/Elodie, Whipple Observatory, ESO/CES and McDonald. Approximately 200 exoplanets have been found with periods between a few days and a few years and with $M \sin i$ spanning 0.1–10 $M_J$. Since there is no accepted theoretical model for the planet mass, $M$, or semimajor axis, $a$, distributions, we adopt a simple power law distribution: d$N$/d$M \propto M^{\alpha}$ and d$N$/d$a \propto a^{\beta}$.

Figure 5 shows the results of a Monte Carlo experiment using GPI with ideal apodization, Fried parameter, $r_0$ = 100 cm at the observing wavelength of 1.6 $\mu$m, operating at 2.5 kHz. In this example the planet population is parameterized by a planetary mass spectrum with $\alpha$ = -1, a semi major axis distribution with $\beta$ = -1/2 between 0.1 < $a$/AU < 50, and a star formation history such that the age of the disk is 10 Gyr and the mean age of stars in solar neighborhood is 5 Gyr. For this simple plot, we assume no suppression of residual speckles, and therefore speckle noise always dominates. When this is true, it is a good approximation to assume a universal contrast curve for all targets, which means that the results can be easily visualized in a contrast versus angular separation plot.

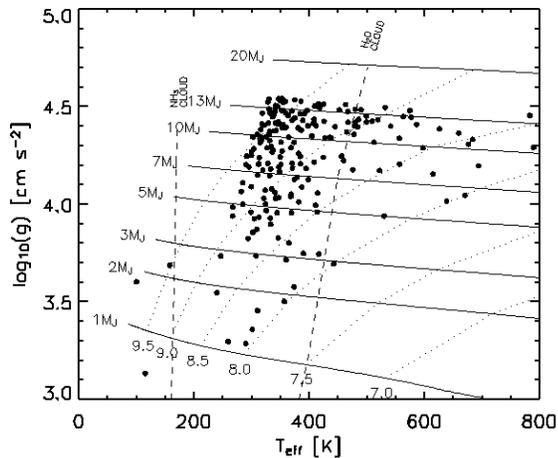

Figure 7: Exoplanets from the simulated GPI field star survey. Solid lines show the evolution of 1–20 $M_J$ exoplanets. Dotted lines are isochrones labeled in $\log_{10}(Gyr)$. Detected planets are filled circles. The population straddles the $H_2O$ cloud condensation line at about 400 K (dashed), and a few objects lie below the $NH_3$ condensation curve (dashed). *The only confirmed astronomical object that lies on this plot is Jupiter,* with $T_{eff}$ = 120 and $\log_{10} g$ = 3.4.

This simple system can detect about 5% of the planets in a survey of field stars. Preliminary calculations of this type were first used to argue for the scientific utility of GPI because the exoplanet detection rate is comparable to that delivered by Doppler searches. The GPI IFU will be used to suppress speckles, but the results are not so easily visualized—in subsequent simulations (e.g., Figure 7) all significant noise sources, e.g., photon shot noise, background, flat field noise, detector read noise and dark current, are fully treated.

High fidelity calculations using the full sensitivity of the system predict that in a survey of approximately 3800 field stars with $I < 8$ mag. we would detect over 200 planets. Figure 7 shows two ways to visualize how GPI-detected planets explore the potential discovery phase space. Figure 6 compares the catalog of Doppler planets, a hypothetical astrometric survey and the GPI exoplanets.

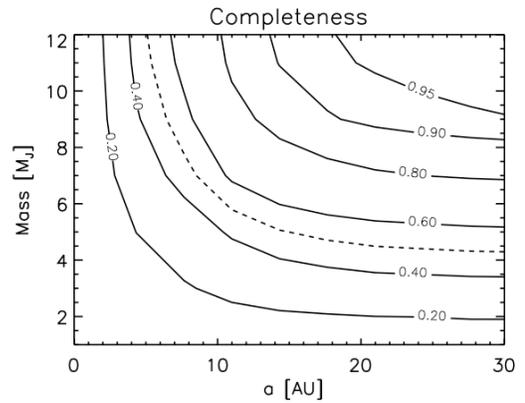

Figure 8: Exoplanet detection completeness contours for an age-selected (< 2 Gyr) GPI survey of stars in the solar neighborhood as a function of semimajor axis and planetary mass. GPI completeness is good (>50%) for 5 $M_J$ planets beyond 10 AU. Massive planets (12 $M_J$) are found with similar completeness as close as 5 AU. Approximately 500 such stars are accessible to GPI ($I < 8$ mag.) in the solar neighborhood and over 100 exoplanets should be detectable.

This comparison confirms that GPI is complementary to indirect methods. GPI can probe the outer regions of solar systems and will answer our key science questions related to planet formation and migration. Figure 7 plots the GPI exoplanets atmospheric properties on an effective temperature, log ($g$) diagram. The youngest detected exoplanets have ages of a few hundred Myr, representing the youngest systems in the field-star sample. The youngest objects sample masses as low as 1 $M_J$. With increasing age, the mean detected planet mass increases. The median exoplanet age is 1.3 Gyr, the oldest have ages of ~ 5 Gyr. About 70% of the planets are cooler than that water cloud condensation line, and three lie to the left of the ammonia cloud condensation curve. The properties of young planets (< 100 Myr) depend on initial conditions, and have been excluded from these Monte Carlo simulations. It is worth emphasizing that at GPI's contrast levels we do not require super-luminous "hot start" planets but can detect planets formed through low-entropy core-accretion.

Important complements to broad field-star surveys are targeted surveys of young clusters and associations or age-selected field stars, where the mean GPI exoplanet detection rate can approach 50%. Figure 7 (right) shows that the median detected exoplanet age in the field survey is 1.3 Gyr. Old stars can be winnowed using a chromospheric activity indicator such as Ca H & K. Employing a selection criterion that eliminates stars older than 2 Gyr boosts the planet detection rate by a factor of four. The completeness of such a survey is represented in Figure 8. About 500 stars in the solar neighborhood would be selected by such a survey, which should yield over 100 exoplanets. This suggests that a preliminary estimate of the semimajor axis and planetary mass distributions could be made in 15–30 nights of telescope time. The Gemini Observatory plans to devote 100-200 nights to GPI campaigns, therefore a number of comprehensive exoplanet surveys will be possible to accumulate statistics relevant to planet formation and to study exoplanet atmospheres.

GPI also enables a broad range of adjunct science. Most relevant to the ExoPTF is sensitive detection and characterization of planetary debris disks. Unseen planets below GPI's detection threshold may gravitationally sculpt debris disks, and therefore the ability to detect disks with optical depths as small as $8 \times 10^{-5}$, or approximately one fiftieth of well-known systems such as beta Pic or AU Mic, is a key feature. Other science missions include high contrast imaging of minor planets and icy moons in the solar system, studies of main sequence binaries (GPI can see an M dwarf next to an O star), and investigation of mass loss from evolved stars.

## Conclusions & Recommendations to the ExoPTF

Outer planets ($a > 10$ AU) take more than thirty years to complete one revolution. These planets will remain undetected in the first-generation Doppler surveys until about 2030. In contrast, direct detection of exoplanets is feasible now using advanced AO techniques. Within three years the GPI instrument will begin to survey for exoplanets beyond 10 AU. GPI will give our first glimpse of the outer regions of exoplanetary systems, where novel planet formation pathways may operate, the counterparts of inward planetary migration may reside, and Kuiper belt analogs may be discovered. Around selected younger targets, GPI will have significant sensitivity in the 5-10 AU range. From low-resolution spectra of the atmospheres of exoplanets GPI will also yield effective temperatures, gravity and composition. As static wavefront errors set the fundamental noise floor for direct detection, whether on the ground or in space, GPI is a unique pathfinder for future NASA missions. The GPI program therefore represents the next major step in direct planet detection and should remain a priority of the Gemini Observatory.

The Gemini Observatory has recognized that the full scientific potential of GPI can only be realized if observing with this instrument is coordinated into major campaigns. Substantial resources are needed immediately if the US astronomical community is to compete effectively to conduct these hundred-night observing programs. Foundation science includes the assembly of catalogs of adolescent (0.1–1 Gyr) stars, which are the most fertile hunting grounds for self-luminous exoplanets. Efficient use of observing time will require careful experimental design, and resources will be needed to reduce and analyze GPI data for reliable exoplanet discovery and characterization. Support will also be necessary to prepare for follow up observations of candidates, including building a strong theoretical basis for understanding exoplanets atmospheres in the temperature range of 200–600 K.

## References

Angel, J. R. P. 1994, *Nature* 368, 203.
Boss, A. P. 2006, ApJ, 644L, 79.
Burrows, A., Sudarsky, D., & Hubeny, I. 2004, ApJ, 609, 407.
Graham, J. R. 2002, BAAS, 201 2102
Grillmair, C. J., et al. 2007, ApJ, 658L, 115.
Macintosh B. et al. 2006, SPIE, 6272, 18
Marois, C., Doyon, R., Racine, R., & Nadeau, D. 2004, PASP, 112, 91
Marois, C. et al. 2006, SPIE, 6269, 114

Marley, M. et al. 2007, ApJ, 655, 541.
Morzinski, K., M., et al. 2006, SPIE, 6272, 62
Poyneer et al. 2006, 6272, 50
Richardson, L. J., et al. 2007, *Nature*, 445, 892.
Sivaramakrishnan, A., & Lloyd, J. P. 2005, ApJ, 633, 528
Soummer, R. 2005, ApJ 618L, 161
Wallace, J. K., et al. 2006, SPIE, 6273, 22